\documentclass[12pt]{article}

\usepackage{paper2e}
\usepackage{mydefs2e}
\usepackage{xspace}
\usepackage{graphicx} 


\newcommand{\LQ}{\Lambda_{\rm had}}
\newcommand{\aQ}{\alpha_{\rm QCD}}
\newcommand{\Rh}{R_{\rm had}}
\newcommand{\Tf}{T_{B}}

\begin{document}

\begin{titlepage}
\preprint{UMD-PP-06-058\\ UFIFT-HEP-06-18}

\title{The Relic Abundance of\\\medskip
Long-lived Heavy Colored Particles}

\author{Junhai Kang,$^{\rm a}$\ \ Markus A. Luty,$^{\rm a}$
\ \ Salah Nasri$\,^{\rm a,b}$}

\address{$^{\rm a}$Physics Department, University of Maryland\\
College Park, Maryland 20742}

\address{$^{\rm b}$Department of Physics, University of Florida\\
Gainesville, Florida 32611}

\begin{abstract}
Long-lived colored particles with masses $m \gsim 200\GeV$ are allowed
by current accelerator searches, and are predicted by a number
of scenarios for physics beyond the standard model.
We argue that such ``heavy partons'' effectively have a 
geometrical cross section (of order $10$~mb) for annihilation
at temperatures below the QCD deconfinement transition.
The annihilation process involves the formation of an intermediate
bound state of two heavy partons with large orbital angular momentum.
The bound state subsequently decays by losing energy and
angular momentum to photon or pion emission,
followed by annihilation of the heavy partons.
This decay occurs before nucleosynthesis for
$m \lsim 10^{11}\GeV$ for electrically charged partons and
$m \lsim \mbox{TeV}$ for electrically neutral partons.
This implies
that heavy parton lifetimes as long as $10^{14}$~sec are
allowed even for heavy partons with $m \sim \mbox{TeV}$
decaying to photons or hadrons with significant branching fraction.
\end{abstract}

\end{titlepage}

\noindent
The phenomenology and cosmology of heavy ($m \gg \hbox{\rm GeV}$)
long-lived colored particles
has received renewed attention recently because of the proposal of
``split supersymmetry'' \cite{splitSUSY}.
Another possible motivation is having a long lived gluino
\cite{raby} or squark \cite{squarkLSP}
as the next-to lightest superpartner in
weak scale supersymmetry.
More generally, it is a phenomenologicall interesting possibility that
such particles could exist.
For $m \gsim 200\GeV$ such particles are consistent with current
collider bounds, and if $m \lsim 2\TeV$ they will be accesible
at LHC \cite{colliders}.

This note concerns the cosmology of such particles,
which we refer to generically as ``heavy partons,'' since they are
constitutents of exotic long-lived hadrons at low energies.
This has been studied by a number
of authors, and there is some controversy about the
the correct relic abundance, even at the level of the order of magnitude
\cite{relic}.
The disagreement is over the extent to which the strong interactions
enhance the annihilation cross section in the early universe over
the perturbative value.
At temperatures below the deconfinement temperature $T_{\rm c} \simeq 180\MeV$,
the heavy partons are confined
inside hadrons. 
Just as in heavy quark effective theory \cite{HQET}
it is useful to picture these hadrons as consisting of a heavy parton
surrounded by QCD ``brown muck'' with a radius of order
$\Rh \sim \hbox{\rm GeV}^{-1}$ (see Fig.~1).
If the heavy parton is a color triplet,
the brown muck will involve at least one light quark, while if the
heavy parton is a color octet, it may involve just gluons.
Our arguments will not depend on the details of the
brown muck.

\begin{figure}[tb]
\begin{center}
\centerline{\includegraphics{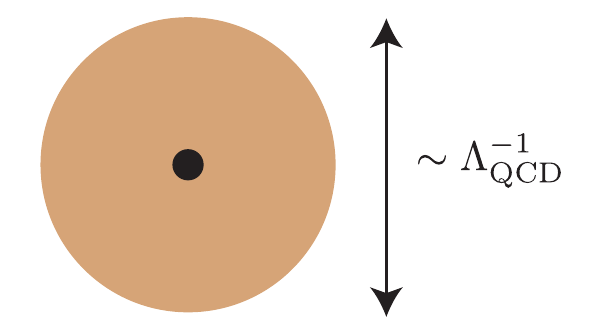}}
\begin{minipage}[t]{5in}%
\caption{A schematic representation of a hadron made
of a heavy parton and surrounded by QCD ``brown muck.''}
\end{minipage}
\end{center}
\end{figure}

We therefore consider two heavy hadrons in the early universe at
temperatures below the QCD phase transition.
It is clear that the strong interactions give a 
geometrical cross section (of order $\Rh^2$) for
the heavy hadrons to \emph{interact}\/, but this only means that the
brown muck of one hadron interacts with that of the other.
In order for the partons to \emph{annihilate}\/, the wavefunctions
of the heavy partons
themselves must overlap, so the direct annihilation cross section
is proportional to $m^{-2}$.

\begin{figure}[tb]
\begin{center}
\centerline{\includegraphics{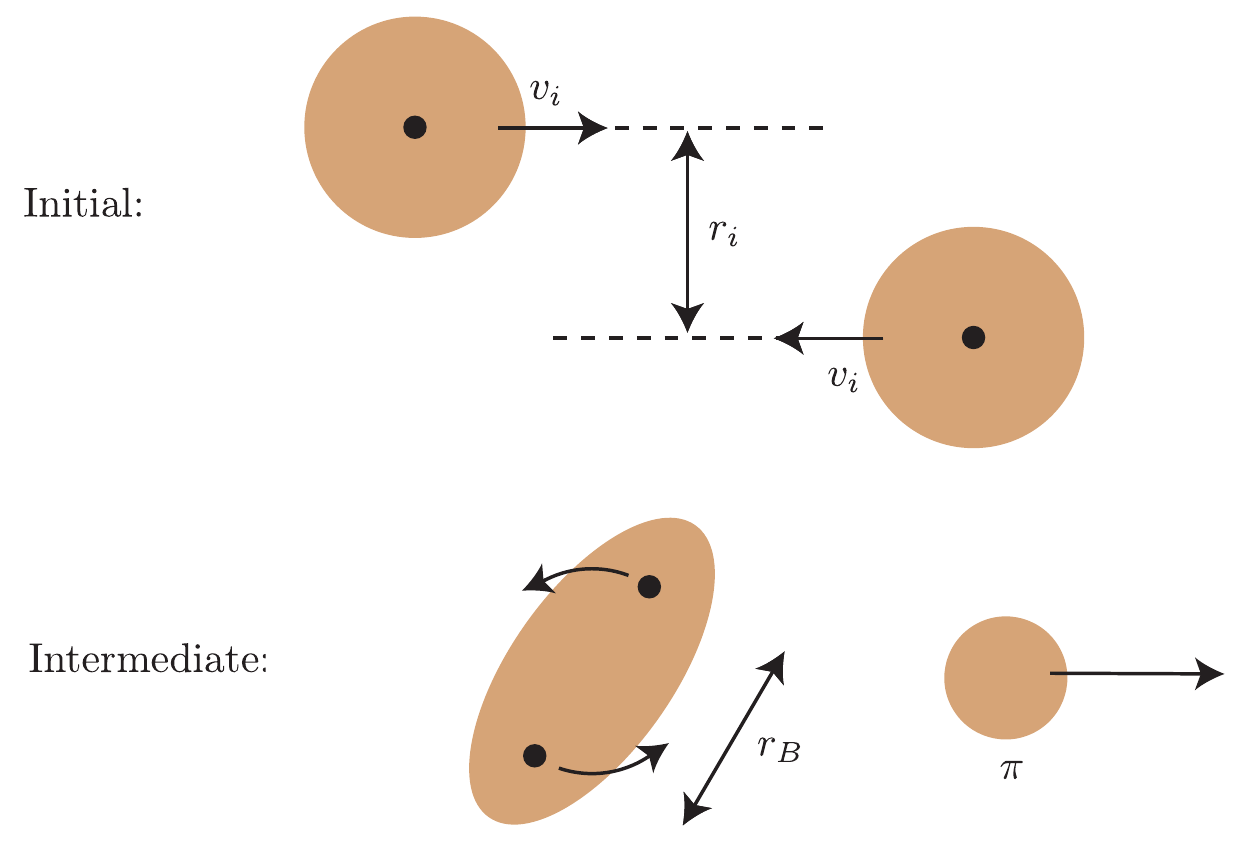}}
\begin{minipage}[t]{5in}%
\caption{The formation of a highly excited bound state.}
\end{minipage}
\end{center}
\end{figure}

We will argue that the strong interactions nonetheless
give rise to an effective geometrical cross section for
annihilation at temperatures below the QCD phase transition. 
The first stage of this process is illustrated in Fig.~2.
This is a capture process in which two heavy hadrons interact to
form a bound state, with the energy carried off \eg\ by a pion.
As we will see, this bound state has large orbital angular momentum
($L \sim 10$ for $m \sim \hbox{\rm TeV}$)
and therefore does not decay promptly.
In the second stage, the bound state loses energy and angular
momentum, \eg\ by emitting pions and/or photons.
Bound states with binding energy larger than the temperature
$T$ survive collisions with particles in the thermal bath,
and eventually decay through annihilation.
The net result is that the heavy partons effectively have an enhanced
cross section for annihilation given by the capture cross section.
(This is similar to the scenario originally described in
\Ref{splitSUSY} for stable gluinos.)
The phenomonological implications of this result will be discussed
at the end of this note.

We now explain why the capture process is unsuppressed.
The first point is the existence of the bound state.
The potential between two heavy partons can be written schematically
as a sum of a short-distance Coulomb interaction and an attractive
linear term representing the effects of confinement:
\beq
V(r) \sim \frac{C \aQ}{r} - \si r.
\eeq
Here $\si \sim \LQ^2$ is the string tension,
and $C$ is a group theory factor
that depends on the color representation of the heavy partons.
The color Coulomb force is always attractive in the color singlet
channel (\ie\ $C < 0$).
We also expect the long-range part of the potential to be
attractive in the color singlet channel,
since it is responsible for color confinement.
The energy spectrum of the system of two heavy partons therefore looks
as shown schematically in Fig.~3.
The low-lying states are Coulombic, with energy splittings of
order $\aQ^2 m \gg \LQ \sim \hbox{\rm GeV}$, while the states near the
continuum threshold are dominated by the linear term.
Spin-dependent interactions are suppressed by $1/m$, so
spin excitations are small and we do not expect them to play
an important role in the process we are considering.
The rotational excitations are very important for determining
the properties of the intermediate bound state.
The minimum radius for a given angular momentum is determined by
the circular orbits.
In the regime where the linear term dominates, we have
\beq[rbound]
r_{\rm min} \sim \left( \frac{L^2}{\si m} \right)^{1/3}.
\eeq
The largest angular momentum occurs for $r \sim \Rh$.
At larger radii, the string confining the heavy partons will
break and the spectrum becomes a continuum of two-particle states.%
\footnote{String breaking is suppressed for large $N_{\rm c}$.
We will not keep track of factors of $N_{\rm c}$ in this paper.}
The largest angular momentum of a bound state is therefore
\beq[Lmax]
L_{\rm max} \sim \left( \frac{m}{\LQ} \right)^{1/2}
\sim 30 \left( \frac{m}{\rm TeV} \right)^{1/2}.
\eeq
This estimate is consistent with the fact that there are
expected to be $L = 3$ states in the the $\Upsilon$ system
below the $B$-$\bar{B}$ threshold \cite{upsilon}.
This gives
$L_{\rm max} \sim 3 ( m / m_b)^{1/2}
\sim 40$ for $m \sim \hbox{\rm TeV}$.

\begin{figure}[tb]
\begin{center}
\centerline{\includegraphics{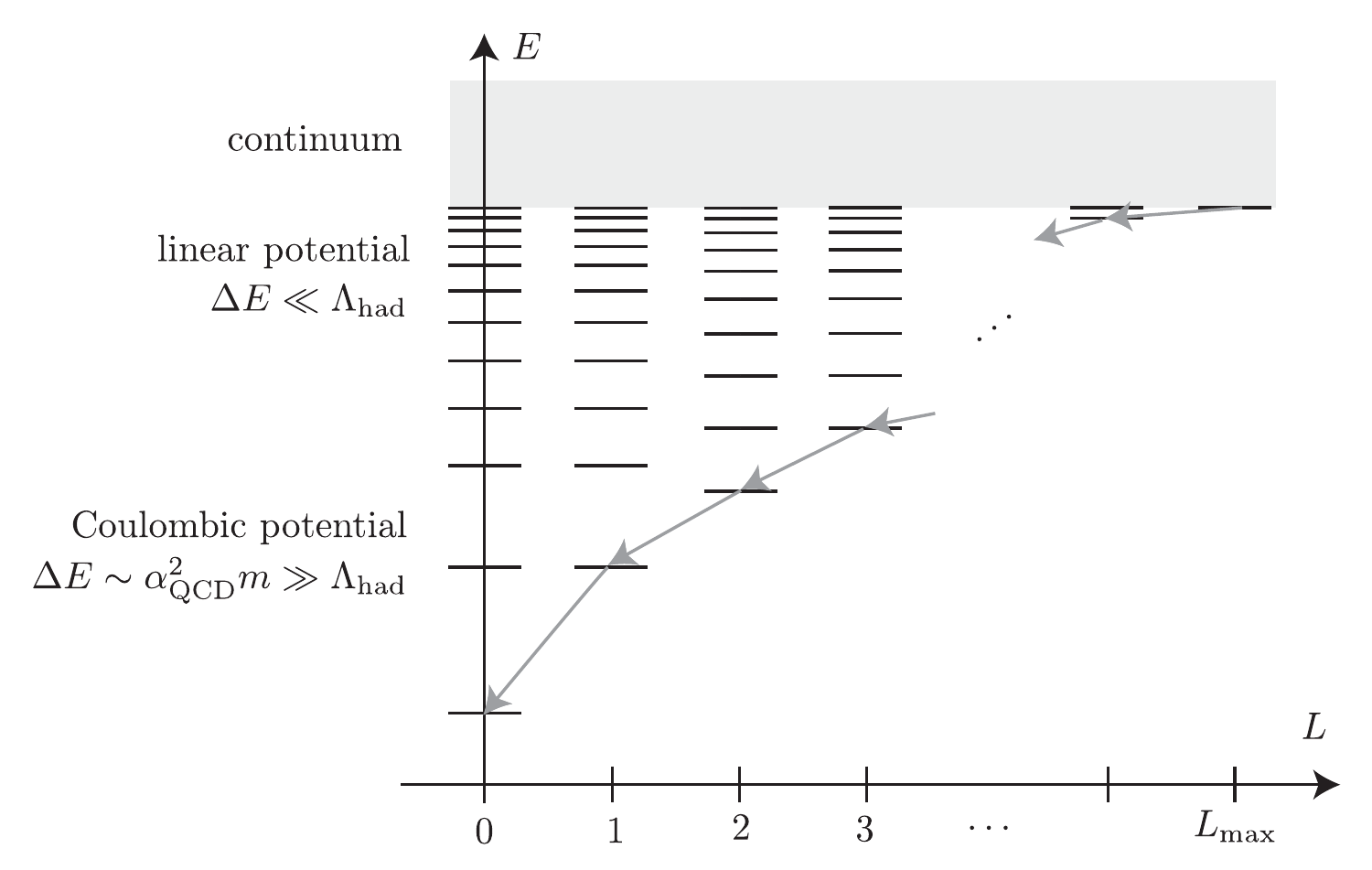}}
\begin{minipage}[t]{5in}%
\caption{Spectrum of the system consisting of two heavy partons.
(Spin excitations are not shown.)
Bound states with large angular momentum will decay
through a series of $\De L = 1$ transitions, as shown.}
\end{minipage}
\end{center}
\end{figure}

We now consider the process depicted in Fig.~2, taking place at
temperatures $T \lsim T_{\rm c}$.
We are interested in impact parameters $r_i \lsim \LQ^{-1}$, so that
the brown muck clouds overlap, and therefore interact strongly.
(We neglect the possibility that pion exchange gives a longer range
attractive interaction.)
To understand the possible transition to a bound state, 
we must understand what quantities are conserved in this transition,
taking into account the special kinematics of the situation.
Unlike a weakly interacting relic, a colored particle will be in kinetic
equilibrium with gluons at the deconfinement temperature.
The initial velocity is therefore determined by the temperature to be
\beq
v_i \sim \left( \frac{\Tf}{m} \right)^{1/2},
\eeq
where $\Tf$ is the temperature where the bound state is formed.
Note that the momentum is large compared to the inverse
range of the interaction
\beq
p_i \sim m v_i \sim (m \Tf)^{1/2} \gg \frac{1}{\Rh},
\eeq
so the scattering process is not dominated by $s$-wave scattering,
and we expect higher partial waves to be important.
On the other hand, the
velocity is slow enough that the forces exerted by the
brown muck can change the velocity significantly:
\beq[vchange]
\frac{\De v}{v_i} \sim \frac{a \De t}{v_i}
\sim \frac{(\LQ^2 / m) (\Rh/v_i)}{v_i}
\sim \frac{\LQ}{\Tf} \gsim 1.
\eeq
We see that the velocity is not conserved in the collision,
even though the partons are heavy.

Energy and angular momentum must of course be conserved in
the collision.
Energy conservation is satisfied by the emission of a pion
(or perhaps several) that carries away the binding energy.
The pion can also carry away some angular momentum, but only
$\De L \sim \hbox{\rm few}$.
The typical initial angular momentum is
\beq
L_i \sim m v_i r_i \sim  (m \Tf)^{1/2} \Rh
\sim 10 \left( \frac{m}{\TeV} \right)^{1/2}
\left( \frac{\Tf}{T_{\rm c}} \right)^{1/2}.
\eeq
The initial angular momentum is large, but note that $L_i$
becomes smaller than $L_{\rm max}$ for $\Tf < T_{\rm c}$.
This means that the binding energy is
\beq
B = E_{\rm max} - E_f \sim \left( \frac{\si^2 L_{\rm max}^2}{m} \right)^{1/3}
- \left( \frac{\si^2 L_{i}^2}{m} \right)^{1/3}
\sim \left( \frac{\si^2 L_{\rm max}^2}{m} \right)^{1/3}
\sim \LQ.
\eeq
The initial kinetic energy is only $T_{\rm c} \sim m_\pi$, but the
binding energy is sufficiently large to produce a pion.

It is also easy to see that the transverse distance between the heavy
partons is allowed to change significantly during the collision.
The forces of the brown muck give
\beq
\De r_\perp \sim  a_\perp \De t^2  \sim 
\frac{\LQ^2}{m} \left( \frac{\Rh}{v} \right)^2 
\sim \frac{1}{\Tf} \gsim \Rh,
\eeq
where $\perp$ refers to the component perpendicular to
the line connecting the heavy partons.
Conservation of angular momentum gives
\beq
\frac{\De r_\perp}{r} \sim \frac{\De v_\perp}{v} 
\gsim 1.
\eeq

Based on the considerations above, we argue that
the cross section for the formation of the bound
state is geometrical, \ie
\beq
\si_{\rm form} \sim \pi \Rh^2.
\eeq
The reason is simply that there is no symmetry or kinematic
factor that suppresses the transition, so it should proceed at the
going rate for strong interactions.
In fact, it is possible that longer range forces due to
pion exchange
increase the cross section, a possibility that we will neglect
here.
As we have seen, the transition can occur for impact parameters
of order $\Rh$ without violating any exact or approximate
conservation law.
Furthermore, the energy of the emitted pion(s) is of order
$\LQ$, so there is no low-energy suppression.

It is important to keep in mind that the geometrical cross section
gives the rate for the bound state to \emph{form}\/, and we must
consider the subsequent evolution of the bound state to see whether
it enhances the \emph{annihilation} of the heavy partons themselves.
The first issue is whether bound states of the kind we are discussing
can survive in the thermal bath long enough so that it can radiate away
energy and angular momentum and eventually annihilate.
The biggest threat to their survival is collisions with photons,
since pions have low number density for $T < m_\pi \sim T_{\rm c}$.
States with binding energy $B > T$ cannot be destroyed efficiently,
since the probability for finding a photon in the thermal bath
with energy $B$ is suppressed by $e^{-B/T}$,
which drops rapidly for $T < B$.
(This rate will have additional suppression if the heavy partons
and the brown muck are electrically neutral.)
To find the precise value of the temperature below which destruction
becomes inefficient requires a detailed hadronic model, but it is
of order $T_{\rm c}$.

One might also worry about successive collisions with photons gradually
increasing the energy of the bound state until it becomes unbound.
The dominant process is inverse decay $B\ga \to B'$,
where $B$ and $B'$ are bound states.
However, these merely establish an equlibrium distribution
of excitations (in which bound states are more numerous)
on a time scale given by the rate for the decay $B' \to B\ga$.

We now consider the decay of the bound state.
We consider first the case of electrically charged heavy
partons, which can decay by the familiar process of photon radiation.
The bound state decays by first radiating away energy and angular
momentum to get to a bound state with $L \sim 1$,
which then decays by annihilation into quarks or gluons.
We first estimate the decay rate to an $L \sim 1$ state.
The linear term dominates the force on a heavy parton
for $r \gsim R_{\rm c}$, where
\beq[Rc]
R_{\rm c} \sim \left( \frac{\aQ}{\si} \right)^{1/2}.
\eeq
This corresponds to a binding energy
\beq
E_{\rm c} \sim \frac{\aQ}{R_{\rm c}} \sim (\aQ \si)^{1/2}.
\eeq
The time to lose energy of order $E_{\rm c}$ can be estimated by
using the Larmor formula $\dot{E} \sim \al a^2$, where $a \sim \si/m$
is the acceleration in the linear potential.
Since the acceleration is constant, we have
\beq[lifetimelinear]
\tau(\De E \sim -E_{\rm c}) \sim \frac{E_{\rm c}} {\dot{E}}
\sim \frac{\aQ^{1/2} m^2}{\al \LQ^3}.
\eeq
The time to lose the remaining energy can again be estimated by
using the Larmor formula, with $a$ determined from the Coulomb
potential.
It is easy to check that this is dominated by the energy loss
for binding energies of order $E_{\rm c}$, and we get the same
estimate \Eq{lifetimelinear} for this decay time.
The subsequent annihilation is much more rapid, as can be seen for
example from the formula for the annihilation rate for an $L = 0$
state:
\beq
\Ga_{\rm annihilation} \sim \frac{4\pi \aQ^2}{m^2} | \psi(0)|^2,
\eeq
where $\psi(0)$ is the radial wavefunction of the bound state
evaluated at the origin.
Since $\psi(0) \sim (\aQ m)^{3/2}$ for the ground state, we have
$\Ga_{\rm annihilation} \sim 4\pi \aQ^5 m$.
We conclude that the decay rate for the bound state is given by
\Eq{lifetimelinear}.
This decay occurs before nucleosynthesis 
($\tau \lsim 1~\mbox{sec}$) for $m \lsim 10^{11}\GeV$,
where we have used $\LQ \sim \GeV$.
For larger masses, the late decays to photons
will affect nucleosythesis, and the model is ruled out
\cite{nucleodecaybounds}.

We now discuss the decays of the bound state in the case where the
heavy partons are electrically neutral.
We assume that the brown muck is also electrically neutral.
The energy can then
be carried away by either photons or pions.
The photon rate is a loop effect, and is suppressed by the small electromagnetic
coupling, but the pion rate is potentially kinematically suppressed by the
pion mass, which can be larger than the energy differences between states
with $\De L \sim 1$ for large $m$.
We will confine our attention to the two photon decays, since we will see
that they are sufficiently rapid for the most interesting range of $m$.

\begin{figure}[tb]
\begin{center}
\centerline{\includegraphics{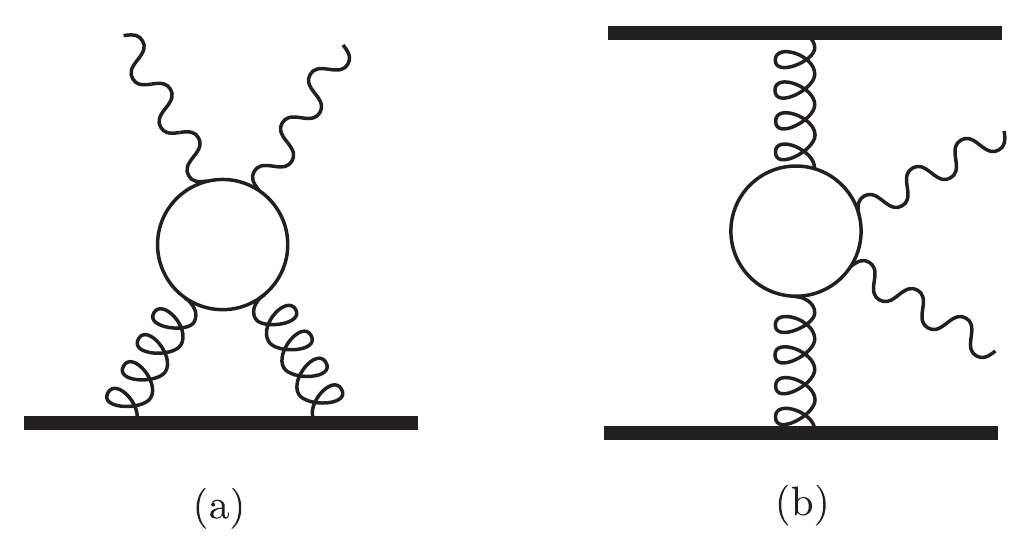}}
\begin{minipage}[t]{5in}%
\caption{Diagrams contributing to the decay
$B' \to B \ga\ga$.
The thick lines are heavy partons and the
thin lines are light quarks.}
\end{minipage}
\end{center}
\end{figure}

Two photon decays occur via diagrams such as the ones in Fig.~4.
Since we are interested in energy and momentum transfers small compared
to $\LQ$, we can parameterize the effects of diagrams such as the
one in Fig.~4a by effective
operators in the heavy parton effective theory:
\beq[HPETphotondecayop]
\!\!\!\!\!\!\!
\scr{L}_{\rm int} \sim \frac{e^2}{\LQ^3}
F^{\mu\rho} F^\nu{}_\rho \psi^\dagger
\left( v_\mu + \frac{i D_\mu}{m} \right)
\left( v_\nu + \frac{i D_\nu}{m} \right) \psi
\sim \frac{e^2}{\LQ^3}
F^{0i} F^{ij} \psi^\dagger \frac{i \partial_j}{m} \psi
+ \cdots
\eeq
Here $\psi$ is a heavy parton field of dimension $\frac 32$
and $v_\mu$ is the 4-velocity of the heavy parton.
We require two photon fields by charge conjugation invariance, and
electromagnetic gauge invariance then forces this to be proportional
to the field strength tensor.
Since we are interested in orbital transitions of the heavy parton, we want a term
with at least one spatial derivative acting on the heavy parton field.
The factor of $1/m$ arises because the heavy parton effective
theory can depend on the 4-velocity
and the residual momentum only in the combination $p_\mu = m v_\mu + k_\mu$
\cite{reparaminv}.
(We ignore the spin indices, which are not relevant for estimating the rate
for orbital transitions.)
Note that derivatives acting on the photon fields give powers of
$E_\ga \sim \De E \ll \LQ$, and so the expansion in terms of derivatives
acting on the photon fields should be valid.
Derivatives acting on the heavy parton fields give powers of $1/r_B$, where
$r_B$ is the size of the bound state.
Since $r_B \gg 1/m$, these should also be suppressed.
There are also 4-parton processes such as the one shown in Fig.~4b that give
rise to 4-parton operators that are nonlocal on the scale $\Rh$.
These cannot be treated as local in the effective field theory we are
considering, and it is more subtle to do the power counting for these
operators.
The operator \Eq{HPETphotondecayop} can be treated using standard
quantum mechanics techniques, and we are confident that it correctly
counts the powers of $m$, $\LQ$, and $E_\ga$ in the process.
We will therefore use it to get at least an upper bound on the
lifetime of the bound state.

Integrating over phase space, we obtain
\beq
\Ga(B' \to B\ga\ga) \sim \frac{\al^2 E_\ga^7}{4\pi m^2 \LQ^6 r_B^2},
\eeq
where $r_B$ is the size of the bound state.
The energy of a bound state in the linear regime is
\beq
E \sim \left( \frac{\si^2 L^2}{m} \right)^{1/3},
\eeq
where $E$ goes from $E_{\rm min} \sim \si R_{\rm c}$
(where the potential becomes Coulombic, see \Eq{Rc})
to $E_{\rm max} \sim \LQ$ (where the system becomes unbound).
In terms of this energy variable, the energy difference
between $\De L = 1$ states is
\beq
\De E \sim \left( \frac{\si^2}{m E} \right)^{1/2}.
\eeq
Using $E_\ga \sim \De E$, we can write the rate of energy loss as
\beq
\dot E \sim \Ga\, \De E \sim \frac{\al^2 \LQ^{14}}{4\pi m^6 E^6}.
\eeq
The lifetime is therefore
\beq
\tau & \sim \myint \,\frac{dE}{\dot{E}}
\sim \int_{E_{\rm c}}^{E_B} dE\,
\frac{4\pi m^6 E^6}{\al^2 \LQ^{14}}
\nonumber
\\
&\sim \frac{4\pi m^6 E_B^7}{\al^2 \LQ^{14}}
\sim \frac{4\pi m^6}{\al^2 \LQ^7}
 \left( \frac{\Tf}{\LQ} \right)^{7/3},
\eeq
where we have used
\beq
E_B \sim \left( \frac{\si^2 L_i}{m} \right)^{1/3}
\sim (\LQ^2 T_B)^{1/3}.
\eeq
Note that the lifetime is dominated by the bound states closest
to threshold, where $\De E \ll \LQ$ and the expansion above is valid.
This decay occurs before nucleosythesis for $m \lsim 2.5\TeV$,
where we use $T_B \sim 200\MeV$ and $\LQ \sim \mbox{GeV}$ 
to obtain the bound.
(Note that since the rate is proportional to $m^6$, the value of $m$
is actually quite well determined.)
It is possible that the decay rate to pions is more rapid, but we 
will not attempt to estimate it here.

Let us comment on the theoretical uncertainties in our analysis above.
There are many numerical factors that we have estimated to be of order 1,
and it is certainly possible that some of our estimates are numerically inaccurate.
However, since we are interested in setting cosmological
\emph{limits}\/, the most
conservative assumptions are those that weaken the limits.
In the analysis above, we have attempted to make ``middle of the road''
estimates for all quantities.
To strengthen the cosmological limits on heavy
stable partons, one would have to demonstrate that the
estimates above are incorrect, taking into account the
large uncertainty in hadronic quantities.
Taking this into account, we believe that the bounds we obtain
are robust.

We now discuss the relic abundance of the heavy partons.
At a temperature of order $T \sim m/30$ perturbative
annihilation of heavy partons due to perturbative QCD gives a
relic abundance
$Y = n_P / s \sim 10^{-14}$ for temperatures $T \lsim m/30$.
Below the QCD phase transition, the second stage of annihilation
described above
further reduces the relic abundance and determines
the final relic abundance.
For $T = \Tf \lsim T_{\rm c}$ the
thermally averaged rate for annihilations
(more precisely, formation of bound
states that later decay) is
\beq
\avg{\si |v|} = \pi R^2 \left(\frac{\Tf}{m} \right)^{1/2},
\eeq
where we expect $R \sim \Rh$.
This reduces the relic abundance until the annihilation rate
drops below the Hubble expansion rate:
$\Ga = n_P \avg{\si |v|} \lsim H \sim g_*^{1/2} T^2 / M_{\rm P}$.
Saturating this inequality gives
a relic abundance of unbound partons of
\beq
Y_P = \frac{n_P}{s}
\sim 10^{-18}
\left( \frac{R}{\mbox{GeV}^{-1}} \right)^{-2}
\left( \frac{\Tf}{180\MeV} \right)^{-3/2}
\left( \frac{m}{\mbox{TeV}} \right)^{1/2},
\eeq
where $s = 2\pi^2 g_* T^3 / 45$ is the entropy density,
and we use $g_* \simeq 10$ just below the QCD phase transition.
We are neglecting entropy production at the QCD phase transition.

This is a very interesting relic abundance for the
phenomenologically relevant mass range $m \lsim \mbox{TeV}$.
The cosmological bounds depend on the
lifetime and decay modes of the heavy partons.
If the lifetime is in the range of $10^2$~sec to $10^6$~sec,
there are bounds
arising from the fact that photon or hadronic decay products
can affect nucleosynthesis \cite{nucleodecaybounds}.
For lifetimes in the range $10^6$~sec to $10^{12}$~sec,
there are bounds coming from the photodissociation of
light elements \cite{nucleodestructionbounds}.
For lifetimes in the range $10^6$~sec to $10^{13}$~sec,
there are bounds from distortions of the cosmic microwave
background \cite{CMBdecaybounds}.
These bounds apply because we
expect the decaying parton to have a significant brancing
ratio into both hadrons and photons.
Remarkably, all of these bounds become ineffective for
$m Y_P \lsim 10^{-14}$~GeV, near the limit of our
estimated relic abundance for the phenomenologically relevant range
of masses $m \lsim \mbox{TeV}$.
The conservative conclusion is therefore that such heavy
partons are not excluded.
For lifetimes in the range $10^{14}$~sec to $10^{18}$~sec, there are
bounds from observations of the diffuse photon background
\cite{diffusedecaybounds}.
These require that the lifetime of the partons is shorter
than about $10^{14}$~sec if there is a significant branching
fraction into photons or hadrons.
Stable partons are ruled out by searches for heavy hydrogen
\cite{heavyHbounds}.
Thus, the reduction in the relic density by several orders
of magnitude over the perturbative prediction lengthens the
allowed lifetime for TeV scale heavy partons
by roughly 12 orders of magnitude!

\section*{Acknowledgements}
We thank P. Bedaque. T. Cohen, T. Okui,
and J. Wacker for discussions.
This work is supported by the National Science Foundation
grant no. PHY-0354401 and the University of Maryland Center for
Particle and String Theory.

\newpage

\end{document}